# Integrity Coded Databases

An Evaluation of Performance, Efficiency, and Practicality

Dan Kondratyuk, Jake Rodden, Elmer Duran
Dept. of Computer Science
Boise State University
Boise, Idaho 83725, USA
dankondratyuk@u.boisestate.edu
jakerodden@u.boisestate.edu
elmerduran@u.boisestate.edu

Jyh-haw Yeh, Archana Nanjundarao
Dept. of Computer Science
Boise State University
Boise, Idaho 83725, USA
jhyeh@boisestate.edu
archanananjundarao@u.boisestate.edu

*Abstract*—In recent years, cloud database storage has become an inexpensive and convenient option for businesses and individuals to store information. While its positive aspects make the cloud extremely attractive for data storage, it is a relatively new area of service, making it vulnerable to cyber-attacks and security breaches. Storing data in a foreign location also requires the owner to relinquish control of their information to system administrators of these online database services. This opens the possibility for malicious, internal attacks on the data that may involve the manipulation, omission, or addition of data. The retention of the data as it was intended to be stored is referred to as the database's integrity. Our research tests a potential solution for maintaining the integrity of these cloud-storage databases by converting the original databases to Integrity Coded Databases (ICDB). ICDBs utilize Integrity Codes: cryptographic codes created alongside the data by a private key that only the data owner has access to. When the database is queried, an integrity code is returned along with the queried information. The owner is then able to verify that the information is correct, complete, and fresh. Consequently, ICDBs also incur performance and memory penalties. In our research, we explore, test, and benchmark ICDBs to determine the costs and benefits of maintaining an ICDB versus a standard database.

## I. Introduction

Cloud database storage is enticing for information storage as it allows customers to pay a flat, recurring rate rather than purchasing the necessary equipment and hiring professionals to create, update, and maintain a local database [10]. The convenience of outsourcing database management to the cloud comes with the potential of malicious attacks however, both from internal and external sources [11]. A customer must trust that the hosting company has taken the proper measures to defend against outside cyber-attacks and also that the company insiders will not attempt to manipulate the data contained within the database [11].

### A. Database Integrity

One approach to ensure database integrity is to utilize an **Integrity Coded Database (ICDB)**. This paper will make extensive references to Boise State University Professor Jyh-haw Yeh's paper on ICDBs: "Protecting Data Freshness and Correctness for Outsourced Databases in Clouds" [3]. It is highly recommended to read the main implementation model in the document, as this paper will provide real-world data on the basic implementation.

By converting a database to an ICDB, the data owner is able to verify that their queried information's integrity is preserved. An ICDB's defining feature is that it contains **Integrity Codes**, cryptographic codes that are stored alongside a unit of data in the database [3]. The integrity codes are generated using a cryptographic function by using database data as an input, and is encrypted with a key that only the database owner has access to. When an ICDB is queried, the queried data is fetched along with the corresponding integrity code. Use of the private key allows the database owner to verify that the information returned by their query matches the expected return information. If either the data or the code is manipulated, the data owner will be able to detect these changes [3].

In order to guarantee integrity, the following criteria should be enforced [3]:

- **Correctness**: Returned data should be original, and not forged.
- **Freshness**: Returned data should be current and not include previously removed data.
- **Completeness**: All matched data should be returned.

ICDBs seek to maintain integrity by avoiding:

- **Data Manipulation**: The alteration of data in the database or in the returned values.
- **Data Omission**: Deletion of data in the database or omission of information in the returned values.
- **Data Addition**: Insertion of data in the database or addition of information in the returned values.
- **Stale Data**: Returning old data or data previously removed from the database.

This document will only be focused on avoiding data manipulation, data addition and stale data. As explained in Yeh's paper [3], it is very difficult to ensure data completeness,

This research was funded by the National Science Foundation under the grant IIS 1461133 and by Boise State University.

and as such the ICDB model implementations in this document will only concentrate on ensuring correctness and freshness.

*B. Database Evaluation*

Although ICDBs are able to detect malicious behavior, they come with a cost. An ICDB will incur memory penalties for storing integrity codes, and performance penalties for creating and returning the codes along with the data. Introducing an additional piece of information to accompany each value in a database increases query time and is detrimental to the overall access and retrieval speed of data. To determine the extent to which the level of efficiency decreases, tests will be performed on three different databases implemented with the MySQL open source database management software [1] and will be testing different benchmarks with the MySQLSlap and MySQL Workbench add-on tools [2]. MySQL is a popular, and provides a wide array of tools and available plugins for additional features.

This research project will be specifically concerned with the following tests:

- **Verification**: Does the introduction of integrity codes reliably detect when data is altered, moved, deleted, or when it is stale?
- **Memory Usage**: How much additional memory is required to store integrity codes?
- **Data Conversion and Retrieval Time**: How quickly can data be converted, retrieved, and verified?

Each of these criteria will be tested by creating a copy of a standard database, converting it to an ICDB, and then running tests on both the DB and the ICDB databases in order to compare results.

This paper has been organized into eight sections. Section II details similar related work, Section 0 presents the encryption process used in generating integrity codes, section IV introduces implementation of ICDB models, section V outlines the specific testing implementation process, section VI reports results of testing, section VII explains possible threats to validity and lastly our conclusions will be reported in section VIII. We have also provided Acknowledgements, References, and an Appendix at the end of this document.

## II. RELATED WORK

*A. Protecting Data Freshness and Correctness for Outsourced Databases in Clouds*

This project's primary focus is to provide results for Dr. Jyh-haw Yeh's paper on integrity coded databases [3]. The paper outlines the implementation of an ICDB using the RSA algorithm to generate integrity codes, one code per field. More explanation on specific implementation will be given in the next section.

*B. Database Security: What Students Need to Know*

Meg Coffin Murray discusses database security and methods by which to improve it [6]. The paper does not delve deeply into database protection from external sources but instead focuses on database security in internal affairs. The discussed method in the paper is to use access control. Different forms of access control allow management to control who has access to what parts of the database. This helps to prevent private data being released and also reduces unwanted changes to the database, by only giving certain users the privilege of reading or writing to particular parts the databases. The second form of security discussed is audit logs. Audit logs store information on what has been accessed and changed and by whom, so that activities are able to be monitored. Both access control management and storage of database modifications are important to our ICDB research as they directly relate to the database owner's ability to ensure the integrity of their information. While access control is relevant, modification logging is invaluable in allowing the database owner to track potential malicious attacks on data.

*C. Database Security Threats and Challenges in Database Forensic: Survey*

[5] explores ways of detecting when a database has been attacked or manipulated, most notably through the use of audit logs and audit databases. The audits save information regarding who has accessed the database and what actions they performed. The logs and databases themselves are also protected to prevent prohibited modifications, removal, or corruption of the information they hold. This allows for quick tracking of individuals who have accessed a database without authorization. Audit databases also give database owners information on what was accessed or changed by the unauthorized individual.

The problem with this method is that it adds more elements to the infrastructure that require protection from unwanted modifications. The audit logs are able to detect changes in the database but must also be able to detect changes in themselves. They must also be able to protect themselves from unwanted modification, or else detecting the changes would prove fruitless.

*D. Hybrid Encryption for Cloud Database Security*

In 2012, Kaur and Bhardwaj from Punjab University conducted research on how to secure data in cloud-stored databases. Their research paper, "Hybrid Encryption for Cloud Database Security" [4] presents a new methodology for encrypting data in the cloud. The proposed solution involves chunking the data in the database and implementing three unique encryption techniques to secure the data. The data is secured by each form of encryption by requiring a different key to access the database information. Because of the difficulty involved in reversing one RSA encryption algorithm, having three layers would be incredibly cumbersome for a hacker to decrypt.

Although a multi-level encryption scheme would bolster the defenses of an ICDB, it would likely come with a dramatic cost in terms of efficiency and overall performance of the database. Because efficiency and performance are of major concern in our research, we have chosen to test the ICDBs with only one of three cryptographic schemes at a time: RSA, Hashing, or AES. These schemes will be discussed in more detail in the following section.

## III. CRYPTOGRAPHIC SCHEMES

Three different cryptographic schemes for verifying database integrity are provided in this document. They are RSA,

Hashing, and AES. Each offers a unique approach to verify data integrity, and each has its own set of strengths and weaknesses.

*A. RSA Encryption*

Most of the emphasis is on RSA public-key encryption. Because we have been given completed modules to test and benchmark different aspects of an ICDB, this implementation will be the primary focus of testing. Although RSA is not the best candidate for practical implementation (see Testing Results), it is still of use due to its mathematical properties.

Regarding integrity code generation, given in [3]:

> The data owner generates all the integrity codes, one for each data item (i.e., each attribute value *T.A(e)*), where *T.A(e)* is the value of an attribute *A* in a table *T* for an entity (i.e., a row) *e*. The integrity code *IC(T.A(e))* for a data item *T.A(e)* is a pair of quantities as described in the following equation.4

$$IC(T.A(e)) = SIG[s \times T.A \times T.A(e) \times T.K(e)] + s \quad (1)$$

> where $\times$ means multiplication, the symbol $+$ means concatenation, s is the unique serial number assigned to the code, *T.A* is the name of the attribute A, *T.K(e)* is the entity *e*'s key value in the table *T*, and *SIG[x]* means the data owner's RSA signature on *x*. The integrity code constructed by the above equation is an unforgeable code that attaches the attribute value *T.A(e)* to its owner entity *e* together. Without knowing the data owner's RSA private key, nobody except the data owner is able to generate a valid signature (or integrity code) on the data [3].

Using this construction, the data owner is able to create an integrity code for each and every field in the table, which contains the attribute, attribute name, primary key, and serial number. To verify correctness, this signature can be regenerated and compared with the stored integrity code. It does not need to be decrypted for verification.

The serial number mentioned earlier is required for freshness. The data owner must keep an **Integrity Code Revocation List (ICRL)**, a list containing the range of serial numbers that are valid. If the data owner updates a table, the serial numbers corresponding to the fields that were removed should be listed in the ICRL file. This serial number will be used in testing for all cryptographic schemes in order to maintain data freshness.

*B. Cyrptographic Hashing*

Cryptographic hashing is the process of taking an input through a function and producing an output such that it is incredibly difficult to determine the input based on the output. It is commonly used in password verification schemes, and can also be used to verify data integrity (e.g., ensuring a file is not corrupted) [12].

A hash function can produce a fixed-length output hash based on an input of any size, making it an ideal candidate for low memory consumption and high performance. A potential downside is that data cannot be recovered from the output hash, which makes verifying individual contents infeasible. If a piece of data does not match its hash, it is not possible to determine what part of the data is incorrect from the hash alone. An integrity code can be generated in the same manner as RSA, except that *SIG[x]* here is a hash function, and the parameters can be concatenated instead of multiplied.

*C. Advanced Encryption Standard (AES)*

AES is a symmetric-key encryption standard that uses a block-cipher for encrypting data. Using a single key for encryption and decryption, the resulting ciphertext can grow in size to accommodate more data [13]. Like RSA, its size must be greater than or equal to the original data size if decryption is desired. This means that using AES (or RSA), the size of an ICDB must be at least double the size of the corresponding standard database.

AES has the advantage of being fast to calculate and relatively low on memory usage. The main advantage is that verification of individual contents is possible, with potential recovery of corrupted data. The next section will show how AES can be used to verify data contents.

## IV. ICDB MODELS

We have implemented two models (methods) of creating integrity codes. They specify what amount data in specific will be stored in an integrity code, which in turn identifies how many integrity codes there will be.

*A. One Code per Field (OCF)*

The OCF model specifies that for every data attribute, there must exist a corresponding integrity code attribute. In other words, for every field containing user data in a table, there must also exist a corresponding integrity code for that data field. The following table illustrates this implementation:

TABLE I.  OCF TABLE STRUCTURE

| *First_Name* | *First_Name_IC* | *Last_Name* | *Last_Name_IC* |
|---|---|---|---|
| George | $IC_1$(George) | Smith | $IC_1$(Smith) |
| Ben | $IC_2$(Ben) | Black | $IC_2$(Black) |
| Bob | $IC_3$(Bob) | Martinez | $IC_3$(Martinez) |

This model allows for a fine-grained control over the data. If a data field does not match its integrity code, the data owner knows that the field is invalid. Data being returned is simply all the data fields along with their integrity codes. If the data owner wants to query for the first name *Ben*, the integrity code *IC(Ben)* will also be returned.

This approach is the one that is adopted from Yeh's RSA model. For testing, both RSA and hashing will use this model.

*B. One Code per Tuple (OCT)*

When testing the OCF implementation, results have indicated that the average ratio between an integrity code and its corresponding data field is extremely high, even for hashing (see Testing Results). The problem is that data fields usually hold only a small amount of data. A tuple with the attribute *First_Name* can contain the field, *George*, which is only 6 characters (48 bits) long. Compare that with an integrity code which are usually recommended to have a minimum of 128 bits to be safe from code forgery. Thus, the integrity codes will most often be much larger than the actual data which they verify.

We have proposed another method of storing integrity codes by using the OCT approach. Instead of generating an integrity code for each and every field, an integrity code will be generated for every tuple (row). This would not only limit the amount of extra fields to store in the database to one per tuple instead of one per field, it would also greatly reduce the integrity code to data ratio. A tuple can contain multiple fields, so the data will be larger, and therefore in many cases will surpass the 128-bit minimum threshold.

To generate an integrity code for an entire tuple, the data owner can concatenate all the data and calculate the signature through a function (denoted $SIG[x]$):

$$IC(T) = SIG[A_1 + A_2 + ... + A_n + s] \qquad (2)$$

where $T$ is the tuple, $+$ is delimited concatenation, $A_1 ... A_n$ are all the attributes in $T$, and $s$ is the unique serial number assigned to the integrity code. Then we generate a signature using some encryption scheme. This signature can be verified to be correct either by decrypting the signature and obtaining the original data (in encryption such as AES), or by encrypting/hashing to this signature and comparing the generated signature to the one stored in the database. For easier retrieval, the serial number can be stored next to the integrity code in another column.

TABLE II. OCT TABLE STRUCTURE

| First_Name | Last_Name | Serial | IC |
|---|---|---|---|
| George | Smith | 1234 | $IC_1$(George, Smith, 1234) |
| Ben | Black | 1235 | $IC_2$(Ben, Black, 1235) |
| Bob | Martinez | 1236 | $IC_3$(Bob, Martinez, 1236) |

Hashing is the best option for memory consumption, since it can produce a fixed-size 128-bit output. However, it is not possible to determine what part of the data is incorrect, only that the tuple is incorrect. Comparing signatures alone will not be enough to verify individual contents, but only whether or not the entire row is correct. Using encryption/decryption, it is possible to decrypt the integrity code and find out which parts of attributes are incorrect. The difference is that decrypting the signature enables checking of individual contents (and hence individual fields) and verifying the correctness of each. Additionally, if the original data cannot be recovered from the integrity code, it is necessary to return all the data in the tuple to regenerate a signature to verify the integrity code. However, if it can be recovered, it is only necessary to return the integrity code, decrypt it, and compare the decrypted data with the returned fields.

AES is the most attractive of the three in this case. Although the integrity code will be at least as large as the data in the tuple, unlike hashing, the integrity code never needs to be recalculated for verification. The integrity code can simply be decrypted and compared against the original data. No additional data needs to be returned, other than the integrity code itself. In short, hashing provides the best memory usage at fixed-length, but requires more data to be returned, decreasing performance. AES may require more memory, but other than the integrity code, does not need to return additional data for verification, and in some cases may require less data returned than hashing.

In this case we have only implemented AES to work with OCT, but hashing is also possible. This document will explore the results of using both OCF and OCT ICDB model implementations. Due to time restrictions, our research only covers three of the six possible ICDB implementations: RSA OCF, hashing OCF, and AES OCT.

V. TESTING PROCESS

A. Hardware and Software Used

All implementation, testing, and benchmarking was performed on Boise State University's Onyx server at onyx.boisestate.edu (See Appendix for specifications). MySQL (MariaDB) was used as the underlying database management system for storing data and executing queries. MySQL Workbench, as well as MySQLSlap were used for performing benchmarks, configured with InnoDB as the database engine and UTF-8 as the character encoding. For database and query conversion, along with query verification, we implemented our own modules using Java SE 1.8.

The three database schemas used for all of our testing are publically available online: *World* (InnoDB) [7], *Sakila* [8], and *Employees* (v1.0.6) [9]. They vary in size from 0.8, 7, and 200 MiB respectively. These schemas are basic examples of databases that one can encounter. For example, *Employees* holds data such as employee information, department numbers, and department locations. Before each test, the databases were returned to their original state to reduce additional variables.

There are ICDB implementations that were tested: RSA OCF, Hashing OCF, and AES OCT. Each implementation will be referred to their respective encryption scheme hence forward (RSA, hashing, AES). RSA uses a 1024-bit (fixed size) output for all integrity codes. Hashing uses PBKDF2WithHmacSHA1 (Java) as its underlying hash function with a 24-byte hash and salt (384 bits total) using 10 hash iterations. AES used ECB (Java) as its implementation. Other implementation modes are possible, but have not been tested.

B. Conversion Process

Our data conversion process is implemented using a MySQL procedure and a few Java Modules. The implementation is given below.

*1) Schema Conversion*

The beginning assumption is that the data owner either has a pre-existing database with data they wish to protect, or the necessary files to create the database. A duplicate schema is created using a schema file. In this implementation, we executed the following query, once for each column in each table in the new database (OCF):

```
ALTER TABLE table_name
ADD COLUMN CONCAT(column_name, '_IC')
TEXT NOT NULL AFTER column_name;
```

where *table_name* is the name of a table, and *column_name* is the name of an attribute within the table (e.g. *First_Name*). This query inserts a new integrity code attribute after a column (e.g. *First_Name_IC*). It is possible to create a dynamic SQL procedure to go through all columns in every table to generate an integrity code attribute alongside every data attribute. OCT is similar, except that only 1-2 (Serial/IC) columns need to be inserted into every table.

*2) Data Conversion*

Once the schema is converted to an ICDB, the data needs to be converted and inserted in the ICDB. Another assumption is that the data is in some kind of file (text in this case) separated by a delimiter. The data owner can do this by executing the following query for each table in the database that contains the data to be converted:

```
SELECT * INTO OUTFILE file_path
FIELDS DELIMITED BY '|'
LINES TERMINATED BY '\n'
FROM table_name;
```

where *file_path* is the pathname of the file that the data will be dumped to. We have created a Java module that will parse the file and automatically generate integrity codes for each field, and create both the ICRL file and the key file to be kept by the data owner. Once that is done, it is a simple case of loading the data files into the new ICDB. It can be executed, once for each table, with the query:

```
LOAD DATA INFILE file_path
REPLACE INTO TABLE table_name
FIELDS TERMINATED BY '|'
LINES TERMINATED BY '\n';
```

This query is used to load all data in existing tables one by one. With that accomplished, the ICDB is ready to accept ICDB queries.

*3) Query Conversion*

A standard query is not sufficient for data verification. The query needs to be converted to return not only the requested data but also the corresponding integrity codes to maintain correctness and freshness. The integrity codes of each field, keys, and attribute names all need to be returned in order to properly regenerate the integrity code.

Dr. Yeh explains OCF query conversion:

Input (an SQL query):

```
SELECT A₁, A₂, ... Aᵣ
FROM T₁, T₂, ... Tₘ
WHERE C₁ and/or ... Cₙ
```

Output (an ICDB query):

```
SELECT A₁, ... Aᵣ , K₁, ... Kₘ, B₁, ... Bₖ,
       IC(A₁), ... IC(Aᵣ), IC(B₁), ... IC(Bₖ)
FROM T₁, T₂, ... Tₘ
WHERE C₁ and/or ... Cₙ
```

where $A_1...A_r$ are attribute names, $K_1...K_m$ are key attribute names for tables $T_1...T_m$ respectively, $B_1...B_k$ are attribute names that appear in the conditions $C_1...C_n$ within the WHERE clause, and finally are those attributes injected for associated integrity codes, one for each of the attributes $A_1...A_r$, $B_1...B_k$ respectively [3].

An example of OCF query conversion is given.

We use two example SQL queries over the "company" database to demonstrate how to apply the above query conversion algorithm.

*Query 1* - "Retrieve the last names of employees who works for department 5": The original SQL query for Query 1 is

```
SELECT lname
FROM employee
WHERE dno = 5;
```

Applying the query conversion algorithm, the corresponding ICDB query is

```
SELECT lname, ssn, dno, IC(lname), IC(dno)
FROM employee
// This employee table is a redefined
// table in the ICDB schema containing
// injected integrity code attributes
WHERE dno = 5; [3]
```

OCT query conversion is as follows.

Input (SQL query):

```
SELECT A₁, A₂, ... Aᵣ
FROM T₁, T₂, ... Tₘ
WHERE C₁ and/or ... Cₙ
```

Output (ICDB query):

```
SELECT A₁, ... Aᵣ , B₁, ... Bₖ, IC₁, ... ICₘ,
FROM T₁, T₂, ... Tₘ
WHERE C₁ and/or ... Cₙ
```

where $A_1...A_r$ are attribute names for tables $T_1...T_m$ respectively, $B_1...B_k$ are attribute names that appear in the conditions $C_1...C_n$ within the WHERE clause, and $IC_1 ... IC_m$ are the integrity codes in tables $T_1...T_m$ respectively. The difference between OCF and OCT is that only the integrity codes in a table column need to be returned, instead of the integrity codes for each and every attribute within a table.

An example of OCT query conversion is given. The query

```
SELECT lname
FROM employee
WHERE dno = 5;
```

will be converted to

```
SELECT lname, dno, IC
FROM employee
WHERE dno = 5;
```

The only additional data required here is *dno* and *IC*, in order to verify that *dno* is indeed 5, and that the *IC* will match against *lname* and *dno*.

## VI. TESTING RESULTS

As mentioned in the Introduction, our criteria for evaluation are: verification, memory usage, and data conversion/retrieval time. The following sections show the results of each criterion evaluated on a standard database and its converted ICDB counterpart. The results are neither exhaustive nor complete, but do provide useful preliminary information about ICDB performance.

*A. Verification*

If an ICDB cannot verify the validity of data, then there is no advantage of using it over a standard database. As such, it is the foremost priority to test whether or not an ICDB will detect various kinds of malicious attacks. The following attacks have been tested.

*1) Forgery Attack*

A forgery attack is an attack that mutates or alters fields in a database. This could involve either the manipulation of the data field itself, or its integrity code.

From the table *Country*, we modified the attribute "Name" of a few entries and were able to detect the changes in each modified field, marked as invalid. Likewise, if we altered the integrity codes for those fields, it would also show up as invalid.

*2) Substitution Attack*

A substitution attack modifies fields by substituting them with other existing fields within the database. This could include copying and replacing a field somewhere else, moving data around, or swapping data from the same row/column.

In the *Country* table, swapping fields with attributes such as "Continent" and "Region" resulted in those swapped entries to be invalid. If both the data field and integrity code were swapped with another data field and integrity code within the same row, we were also able to detect those changes. So data/ICs swapped within the same row/column is able to be detected, along with fields copied over another field.

*3) Old Data Attack*

An old data attack will return correct data (with its integrity code) that was previously stored in the database which has been replaced or deleted.

In the *City* table, removing a row updated the ICRL file accordingly, to flag the corresponding serial numbers as invalid. When trying to return the old data in a SELECT query, verification failed, so those changes were detected.

*4) Tuple Insertion/ Deletion Attack*

A tuple insertion attack introduces new rows in the database. This could be a new piece of data, a row containing data copies from other rows, or a complete duplicate of another row.

Without knowing the keys used for encryption, generating a new tuple would not contain correct integrity codes. We were able to verify this in the *Country* table by inserting new data. Duplicating a tuple is theoretically possible to be undetected by verification, but MySQL prevents tables from having duplicate key values, making duplicate tuples impossible unless the primary key checks were disabled.

A tuple deletion attack removes existing tuples from the database. The verification module was unable to detect deletions from the *City* table. These integrity codes are unable to verify completeness, so tuple deletions will remain undetected.

*B. Memory Usage*

The first benchmark that was performed was measuring memory usage in databases. We analyzed the size of a standard DB, along the size of a converted ICDB with the same basic information, with integrity codes inserted. We also analyzed sizes for some of the tables within each database.

Below, the first chart displays the size increase between DBs and their ICDB counterparts using AES, Hashing, and RSA ICDB schemes respectively.

DATABASE SIZES (MiB)

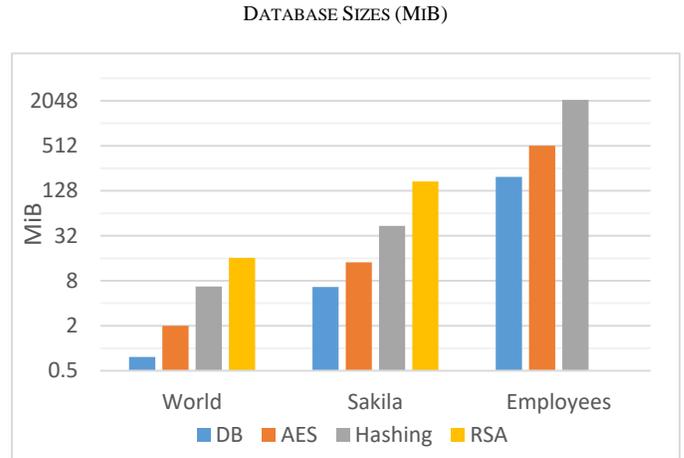

Fig. 1. Database size relationships between three databases. This chart uses a logarithmic base 2 scale to better show database proportions, measured in Mibibytes. RSA data for Employees is not available due to the database conversion taking too long.

Based on the data, the increase in size is quite consistent from database to database. Note that this chart uses a logarithmic (base 2) scale, therefore every gridline is double the size of the previous. On average, database sizes increase by approximately 2.5x, 9x, and 23x through AES, Hashing, and RSA respectively. AES resulted in the least amount of memory used, while RSA resulted in the most.

DATABASE SIZES (MiB)

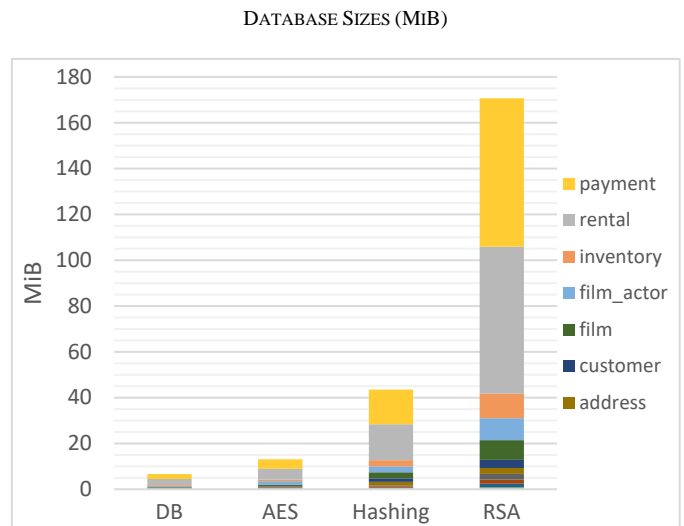

Fig. 2. Database table size relationships in a stacked column layout. Each color represents a table within the database. Larger tables are stacked on top. This chart uses a linear scale measured in Mibibytes.

The second chart shows the individual table sizes within the *Sakila* database, and reveals how much each table has grown from using AES, Hashing, and RSA.

Again, there is a clear progression in size. The tables increased on average 2.6x, 10x, and 25x for AES, Hashing, and RSA respectively. This result is slightly more than the actual database sizes, due to the additional database overhead.

*C. Data Conversion Time*

The next benchmark involved the time required to convert all the data in a standard database to an ICDB. This process involved reading the data file dump attribute for attribute, generating an integrity code based on one or more attributes and copying them in a new ICDB data file. This file can then be read by MySQL and inserted into an ICDB. The time measured in the following chart is the time for converting all data files in a database.

DATABASE CONVERSION TIME (S)

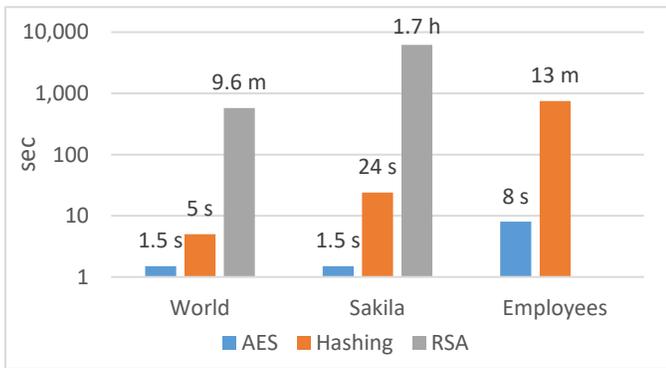

Fig. 3. Database conversion time relationships measured in seconds. Logarithmic base 10 scale. RSA data for Employees is not available due to the database conversion taking too long.

From the chart, AES converted the quickest, hashing was in the middle, while RSA took a long time to convert. Although Employees RSA data is unavailable, we predict that conversion would take at least 18 hours. The results in this section also give some indication as to the increase in time required for ICDB queries, as codes will be need to be regenerated or verified when the database is in use.

*D. Query Execution, Retrieval and Verification Time*

The retrieval speeds of common MySQL queries were analyzed and the results of each query were compared in both the DB and ICDB implementations. Due to difficulty in implementing robust parsing and some time-consuming debugging, we only have a limited set of results regarding query conversion and verification. However, these results do provide some good indication of performance penalties for ICDBs. Each of these queries were tested over 100-1000 iterations and averaged to reduce variability.

The next tests performed involved SELECT * queries to measure the amount of time it takes to return all data in a table. MySQLSlap was used to execute a large number of SELECT * queries on each table in a database, the results from which were averaged. The following chart shows the ratio of the average time increase executing the queries using our different schemes.

SELECT * EFFICIENCY (EXPERIMENTAL / BASELINE)

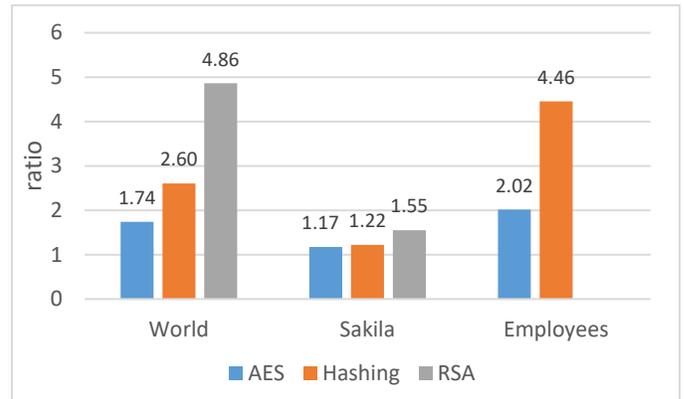

Fig. 4. Database SELECT * time execution ratios, using linear scale. RSA data for Employees is not available due to the database conversion taking too long.

The measurement is the ratio ICDB / DB query execution time. The chart shows that AES executed the queries in the least time, at 1.74x the original time in the *World* database, while RSA executed for the most time, at 4.86x. According to the tables there is a considerable performance penalty for retrieving data, especially for RSA.

Additional results for SELECT, DELETE, and INSERT queries are given below. Due to time constraints, only queries were only tested with RSA on the *World* database. The following chart shows the increase in time required to convert a SELECT query and retrieve the requested data. Queries Q1-Q8 are given in the Appendix. Here, "execution" measures both the carrying out of the query and the data retrieval for the query.

SELECT QUERY CONVERSION, RETRIEVAL, AND VERIFICATION TIME (MS)

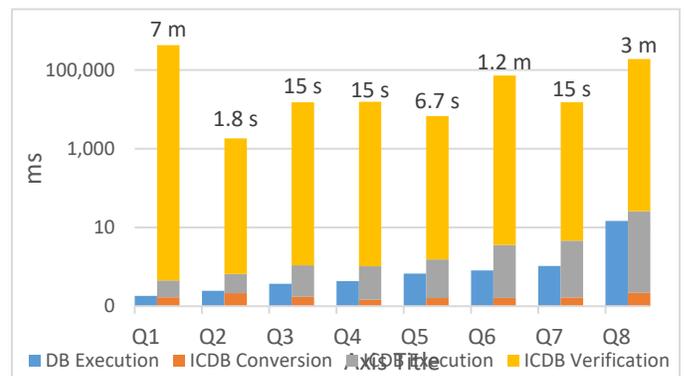

Fig. 5. Database SELECT query execution/retrieval time in blue, and ICDB (RSA) query conversion, execution/retrieval, and verification time in orange, gray, and yellow respectively. Using logarithmic base 10 scale.

This chart compares the execution time of a SELECT query executed on a DB with the same query converted, executed, and verified on an RSA ICDB. The stacked column illustrates the total execution time required to complete all parts. Query conversion is the fastest, and is almost negligible. On average, retrieval takes 2x as long with ICDB data (with ICs) than with DB data. On average, ICDB verification takes 40,000x more time than DB execution/retrieval time. This means that verification dwarfs conversion/execution by taking the vast majority of the time to execute, and will impact performance significantly.

Results for DELETE queries are given in the chart below. Note that before any data is to be deleted, the data should be requested with a select query and be verified. This is to ensure that the data was not modified in any way. However, this is not absolutely necessary, as the modified data would not be used after it is deleted anyway. Once that is done, the ICRL file must be updated to revoke the deleted serial numbers. Queries Q1-Q9 are given in the Appendix.

DELETE QUERY EXECUTION TIME (MS)

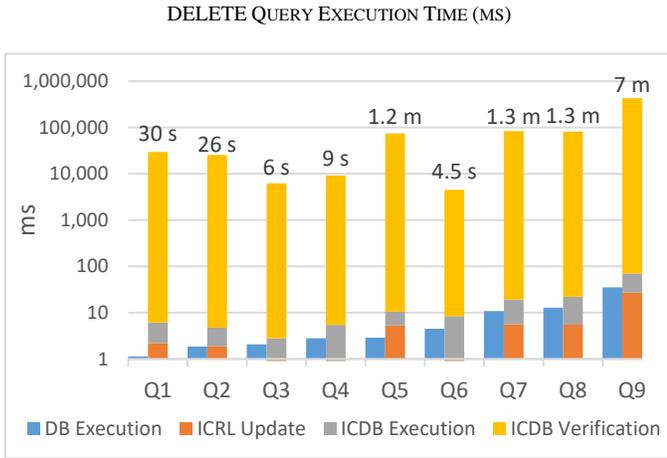

Fig. 6. Database DELETE query execution time in blue, and ICDB (RSA) data verification, query execution, and ICRL update in yellow, gray, and orange respectively. Using logarithmic base 10 scale.

This chart compares the execution time of a DELETE query executed on a DB with the same query verified, executed, and updated on an RSA ICDB. ICDB execution and ICRL updating takes on average 2x more time than standard DB execution. Verification takes the majority of the time, with a 10,000x average increase.

Results for INSERT queries are given below.

INSERT QUERY EXECUTION TIME (MS)

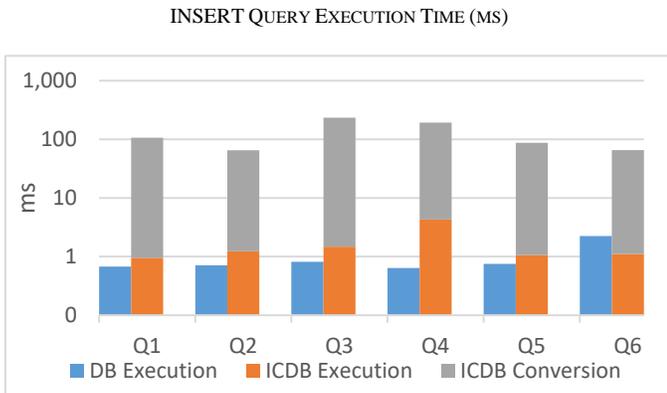

Fig. 7. Database INSERT query execution time in blue, and ICDB (RSA) query execution and conversion in orange and gray respectively. Using logarithmic base 10 scale.

This chart compares the execution of an INSERT statement with the conversion and execution of the ICDB statement. ICDB execution on average took 1.7x more time than DB execution, and ICDB conversion took the bulk of the time, with 130x more time. This result is far less than verification, but still requires a fair amount of additional time to convert each query.

## VII. THREATS TO VALIDITY

Although the benchmarks used for evaluation provide some level of understanding in regard to database performance, they are not exhaustive. There are numerous benchmark standards by which our experiment's databases could have been evaluated. Highly specified tests could be conducted separately, for specific cases which did not fall within the primary focus of this project. Along with this we only ran benchmarks on a MySQL database, and did not test other database options, such as SQLite or Microsoft SQL.

Secondly, the benchmarks were ran a diverse set of databases of differing sizes, however there were limits to the size of the databases we could create and as such we did not test databases with sizes larger than 200 MiB. Results RSA in the *Employees* database do not exist because conversion would have taken 18+ hours. This makes the data slightly inconsistent. Additionally, only three of the six possible ICDB model implementations (RSA OCF, hashing OCF, AES OCT), have been tested. Due to time constraints, RSA OCT, hashing OCT, and AES OCF have not been tested. This does not give a complete picture of how the cryptographic algorithms fare overall in ICDBs. Instead, it offers an indication of possible performance gains using certain implementation models over others.

Thirdly, the ICDB implementations were based on the algorithms explained by Dr. Yeh in his own database research. These may not necessarily be the most effective or sound algorithms by which to generate integrity codes, which is the reason why additional ICDB models are discussed in this document. It is probable that a more robust or efficient algorithm exists which may provide fewer performance penalties.

## VIII. CONCLUSION

From the data, there are many different penalties that are required to run an ICDB over a normal DB. It requires more drive memory to store the ICDB, and more time to return data required for integrity code creation or verification. More memory is required because codes are required to be stored along with each table attribute. More time is required to convert the DB, convert MySQL queries, retrieve data along with their integrity codes, and verify the data, due to more information being required for each query. The goal, then, should be to mitigate this memory and performance penalty as much as possible while not compromising database integrity verification. The simplest approach would be to reduce the sizes of stored integrity codes as much as possible.

An ICDB implementation should allow for secure and certain detection of data. Our implementations were able to detect Forgery, Substitution, and Old Data Attacks and report which fields were invalid. The only type of attack that we tested that was not able to detect changes was the Tuple Deletion Attack, where an attacker could delete existing tuples, or the cloud provider does not return all of the data within the database. In order to detect this attack, enforcement of data completeness is necessary (which is not enforced in this implementation for simplicity, as explained in the Introduction).

The three approaches used are compared here.

A. *RSA (OCF)*
  1) *Alows for decryption and verification of content*
  2) *Massive database size increases*
  3) *Slow to convert*
  4) *Poor query speed*

B. *Hashing (OCF)*
  1) *Alows for verification of content*
  2) *Size is fixed and relatively small*
  3) *Quick to convert*
  4) *Good query speed*

C. *AES (OCT)*
  1) *Alows for decryption and verification of content*
  2) *Size grows with data and is relatively small*
  3) *Very quick to convert*
  4) *Great query speed*

RSA (OCF) is too impractical for large databases. The increase in size (about 25x), along with the slow conversion and verification times indicate that performance will be poor. Hashing (OCF) provides the same functionality as RSA, with a much smaller overhead requirement. The hash could be reduced in size further to increase performance. AES (OCT) provided the fastest and smallest implementation. It is the most practical out of the three, due to its compact representation of codes and quick conversion. It appears that OCT is much more efficient than OCF, due to the smaller amount of data required to return (as explained in section IV, ICDB Models). There are tradeoffs to both approaches.

For specific (non-SELECT *) queries, the only data available is with RSA. One important piece of information is that, in an ICDB, most of the time will be spent verifying the information. Databases can return large amounts of data very quickly, and as such, requesting more information will not dramatically reduce performance. The large bottleneck occurs when the requested data needs to be verified. This is a process that involves looping through every data field/tuple, which can take considerably longer. Performance can be improved significantly by making this operation parallel, as both integrity code generation and verification does not rely on the generation or verification of any other integrity codes. Nevertheless, performing some calculation on the data will still require more resources than simply returning the data. In RSA, the performance penalty is massive, but with other encryption schemes, this could be reduced considerably.

Our data is by no means comprehensive, and there are still many tests that can be performed, along with additional database schemes. The data that is here does provide the extent to which an ICDB could impact performance. Some implementations, such as RSA, could be too costly to the average user. It is necessary to mitigate impact on performance as much as possible while ensuring that malicious attacks can be detected.

From the results of ICDBs that have been tested, using an ICDB in a real-world scenario is feasible, as long as the data owner acknowledges the performance penalties required to maintain data integrity. The results have shown that ICDBs can be light and fast while still maintaining data correctness and freshness. There is much more data to glean from ICDBs than what this document has provided. Going further, it is suggested to explore more types of ICDB models, and additionally, perform more detailed testing.

ACKNOWLEDGEMENTS

We would like to thank Dr. Yeh for his guidance in planning and conducting this research project. Conducting this research would have been impossible without being able to use and study his work on Integrity Coded Databases. We would like to give an additional thank you to our program director, Dr. Xu, for his mentorship and support for this research program. Finally, we give a thank you to Archana Nanjundarao for allowing us to use her Java modules that she wrote as components of Dr. Yeh's research. This work is supported Boise State University, and by US National Science Foundation (NSF) under grant CNS 1461133.

APPENDIX

*A. Hardware Specifications*

All tests were performed on Onyx. Onyx is Boise State University's Linux server provided for students and faculty, accessible through onyx.boisestate.edu. The specifications are provided below.

```
Architecture:          x86_64
CPU op-mode(s):        32-bit, 64-bit
Byte Order:            Little Endian
CPU(s):                24
CPU(s) list:   0-23
Thread(s) per core:    2
Core(s) per socket:    6
Socket(s):             2
NUMA node(s):          2
Vendor ID:             GenuineIntel
CPU family:            6
Model:                 62
Model name:            Intel(R) Xeon(R) CPU E5-2630 v2 @ 2.60GHz
Stepping:              4
CPU MHz:               1453.156
CPU max MHz:           3100.0000
CPU min MHz:           1200.0000
BogoMIPS:              5205.36
Virtualization:        VT-x
L1d cache:             32K
L1i cache:             32K
L2 cache:              256K
L3 cache:              15360K
NUMA node0   CPU(s):   0,2,4,6,8,10,12,14,16,18,20,22
NUMA node1   CPU(s):   1,3,5,7,9,11,13,15,17,19,21,23
```

*B. Queries Used*

The following are the SELECT queries (Q1-Q8) used for the data in Fig. 5. Note that this particular implementation used the "_SVC" suffix instead of "_IC".

DB:

1. SELECT * FROM City;
2. SELECT DISTINCT ID, Name, Population FROM City WHERE `CountryCode`='NLD';
3. SELECT Code, Name, Region FROM Country;
4. SELECT ID, Name, District FROM City WHERE `ID`<'250';
5. SELECT * FROM CountryLanguage WHERE `Language`='English' OR (`Language`='Spanish' AND `IsOfficial`='T');
6. SELECT * FROM Country;
7. SELECT * FROM CountryLanguage;
8. SELECT Country.Name, Country.Continent, Country.Population, City.Name, City.Population FROM Country INNER JOIN City ON Country.Code=City.CountryCode;

ICDB:

1. SELECT ID, ID_SVC, NAME, NAME_SVC, COUNTRYCODE, COUNTRYCODE_SVC,DISTRICT, DISTRICT_SVC, POPULATION, POPULATION_SVC FROM CITY;
2. SELECT DISTINCT ID, ID_SVC, NAME, NAME_SVC, POPULATION, POPULATION_SVC FROM CITY WHERE `COUNTRYCODE`='NLD';
3. SELECT CODE, CODE_SVC, NAME, NAME_SVC, REGION, REGION_SVC FROM COUNTRY;
4. SELECT ID, ID_SVC, NAME, NAME_SVC, DISTRICT, DISTRICT_SVC FROM CITY WHERE `ID`<'250';"
5. SELECT COUNTRYCODE, COUNTRYCODE_SVC, LANGUAGE, LANGUAGE_SVC, ISOFFICIAL, ISOFFICIAL_SVC, PERCENTAGE, PERCENTAGE_SVC FROM COUNTRYLANGUAGE WHERE `LANGUAGE`='English' OR (`LANGUAGE`='Spanish' AND `ISOFFICIAL`='T');"
6. SELECT CODE, CODE_SVC, NAME, NAME_SVC, CONTINENT, CONTINENT_SVC, REGION, REGION_SVC, SURFACEAREA, SURFACEAREA_SVC, INDEPYEAR, INDEPYEAR_SVC, POPULATION, POPULATION_SVC, LIFEEXPECTANCY, LIFEEXPECTANCY_SVC, GNP, GNP_SVC, GNPOLD, GNPOLD_SVC, LOCALNAME, LOCALNAME_SVC, GOVERNMENTFORM, GOVERNMENTFORM_SVC, HEADOFSTATE, HEADOFSTATE_SVC, CAPITAL, CAPITAL_SVC, CODE2, CODE2_SVC FROM COUNTRY;
7. SELECT COUNTRYCODE, COUNTRYCODE_SVC, LANGUAGE, LANGUAGE_SVC, ISOFFICIAL, ISOFFICIAL_SVC, PERCENTAGE, PERCENTAGE_SVC FROM COUNTRYLANGUAGE;
8. SELECT COUNTRY.NAME, COUNTRY.NAME_SVC, COUNTRY.CONTINENT, COUNTRY.CONTINENT_SVC, COUNTRY.POPULATION, COUNTRY.POPULATION_SVC, CITY.NAME, CITY.NAME_SVC, CITY.POPULATION, CITY.POPULATION_SVC, COUNTRY.CODE, CITY.ID FROM COUNTRY INNER JOIN CITY ON COUNTRY.CODE=CITY.COUNTRYCODE;

The following are the DELETE queries (Q1-Q9) used for the data in Fig. 6.

DB:

1. DELETE FROM `Country` WHERE `Continent`='Asia' OR `Continent`='Europe';
2. DELETE FROM `Country` WHERE `IndepYear`<'1950';
3. DELETE FROM `City` WHERE `CountryCode`='ESP';
4. DELETE FROM `CountryLanguage` WHERE `IsOfficial`='T' AND `Percentage`<'50';
5. DELETE FROM `Country`;
6. DELETE FROM `City` WHERE `Population`<'10000';
7. DELETE FROM `CountryLanguage`;
8. DELETE FROM `CountryLanguage` WHERE `CountryCode`!='ABW' AND `CountryCode`!='AFG' AND `CountryCode`!='AGO';
9. DELETE FROM `City`;

ICDB:

1. SELECT CODE, CODE_SVC, NAME, NAME_SVC, CONTINENT, CONTINENT_SVC, REGION, REGION_SVC, SURFACEAREA, SURFACEAREA_SVC, INDEPYEAR, INDEPYEAR_SVC, POPULATION, POPULATION_SVC, LIFEEXPECTANCY, LIFEEXPECTANCY_SVC, GNP, GNP_SVC, GNPOLD, GNPOLD_SVC, LOCALNAME, LOCALNAME_SVC, GOVERNMENTFORM, GOVERNMENTFORM_SVC, HEADOFSTATE, HEADOFSTATE_SVC, CAPITAL, CAPITAL_SVC, CODE2, CODE2_SVC FROM COUNTRY WHERE `CONTINENT`='Asia' OR `CONTINENT`='Europe';
2. SELECT CODE, CODE_SVC, NAME, NAME_SVC, CONTINENT, CONTINENT_SVC, REGION, REGION_SVC, SURFACEAREA, SURFACEAREA_SVC, INDEPYEAR, INDEPYEAR_SVC, POPULATION, POPULATION_SVC, LIFEEXPECTANCY, LIFEEXPECTANCY_SVC, GNP, GNP_SVC, GNPOLD, GNPOLD_SVC, LOCALNAME, LOCALNAME_SVC, GOVERNMENTFORM, GOVERNMENTFORM_SVC, HEADOFSTATE, HEADOFSTATE_SVC, CAPITAL, CAPITAL_SVC, CODE2, CODE2_SVC FROM COUNTRY WHERE `INDEPYEAR`<'1950';
3. SELECT ID, ID_SVC, NAME, NAME_SVC, COUNTRYCODE, COUNTRYCODE_SVC, DISTRICT, DISTRICT_SVC, POPULATION, POPULATION_SVC FROM CITY WHERE `COUNTRYCODE`='ESP';
4. SELECT COUNTRYCODE, COUNTRYCODE_SVC, LANGUAGE, LANGUAGE_SVC, ISOFFICIAL, ISOFFICIAL_SVC, PERCENTAGE, PERCENTAGE_SVC FROM COUNTRYLANGUAGE WHERE `ISOFFICIAL`='T' AND `PERCENTAGE`<'50';
5. SELECT CODE, CODE_SVC, NAME, NAME_SVC, CONTINENT, CONTINENT_SVC, REGION, REGION_SVC, SURFACEAREA, SURFACEAREA_SVC, INDEPYEAR, INDEPYEAR_SVC, POPULATION, POPULATION_SVC, LIFEEXPECTANCY, LIFEEXPECTANCY_SVC, GNP, GNP_SVC, GNPOLD, GNPOLD_SVC, LOCALNAME, LOCALNAME_SVC, GOVERNMENTFORM, GOVERNMENTFORM_SVC, HEADOFSTATE,

```
      HEADOFSTATE_SVC, CAPITAL, CAPITAL_SVC, CODE2, CODE2_SVC
      FROM COUNTRY;
6.    SELECT ID, ID_SVC, NAME, NAME_SVC, COUNTRYCODE,
      COUNTRYCODE_SVC, DISTRICT, DISTRICT_SVC, POPULATION,
      POPULATION_SVC FROM CITY WHERE `POPULATION`<'10000';
7.    SELECT COUNTRYCODE, COUNTRYCODE_SVC, LANGUAGE,
      LANGUAGE_SVC, ISOFFICIAL, ISOFFICIAL_SVC, PERCENTAGE,
      PERCENTAGE_SVC FROM COUNTRYLANGUAGE;
8.    SELECT COUNTRYCODE, COUNTRYCODE_SVC, LANGUAGE,
      LANGUAGE_SVC, ISOFFICIAL, ISOFFICIAL_SVC, PERCENTAGE,
      PERCENTAGE_SVC FROM COUNTRYLANGUAGE WHERE
      `COUNTRYCODE`!='ABW' AND `COUNTRYCODE`!='AFG' AND
      `COUNTRYCODE`!='AGO';
9.    SELECT ID, ID_SVC, NAME, NAME_SVC, COUNTRYCODE,
      COUNTRYCODE_SVC, DISTRICT, DISTRICT_SVC, POPULATION,
      POPULATION_SVC FROM CITY;"
```

The following are the INSERT queries (Q1-Q6) used for the data in Fig. 7.

DB:

```
1.  INSERT INTO `City` (ID, Name, CountryCode, District,
    Population) VALUES ('4080','Boise', 'USA', 'Idaho',
    '123456');
2.  INSERT INTO `City` (ID, Name, CountryCode) VALUES
    ('4080','Boise', 'USA');
3.  INSERT INTO `Country` (Code, Name, Continent, Region,
    SurfaceArea, IndepYear, Population, LifeExpectancy, GNP,
    GNPOld, LocalName VALUES ('TEX', 'Texas', 'North
    America', 'North America', '900000.00', '1900',
    '2790000', '25', '85128', '81111','Texas');
4.  INSERT INTO `Country` (Code, Region, IndepYear,
    LifeExpectancy, GNP, GNPOld, LocalName, GovernmentForm,
    HeadOfState) VALUES ('TEX', 'North America', '1900',
    '25', '85128', '81111','Texas', 'Monarchy','Dhali');
5.  INSERT INTO `CountryLanguage` (CountryCode, Language,
    IsOfficial, Percentage) VALUES
    ('ZZZ','Viatnamese','F','100');
6.  INSERT INTO `CountryLanguage` (CountryCode, Language,
    IsOfficial) VALUES ('ZZZ','Viatnamese','F');
```

ICDB Queries are omitted because they were too large.